\documentstyle[twocolumn,aps,epsfig]{revtex}
\begin{document}
\draft
\title{Baryon Rapidity Loss in Relativistic Au+Au Collisions}

\renewcommand{\thefootnote}{\alph{footnote}}
\author{
	B.B.~Back,$^1$
	R.R.~Betts,$^{1,5}$
	J.~Chang,$^3$
	W.C.~Chang,$^{3}$
	%%\footnote{Institute of Physics, Academica Sinica,
	%%Taipei 11529, Taiwan}
	C.Y.~Chi,$^4$
	Y.Y.~Chu,$^2$
	J.B.~Cumming,$^2$
	J.C.~Dunlop,$^{7}$
	%%\footnote{Yale University, New Haven, CT
	%%06520, USA}
	W.~Eldredge,$^3$ 
	S.Y.~Fung,$^3$ 
	R.~Ganz,$^{5}$
	%%\footnote{Max Planck Institut f\"ur Physik, D-80805
	%%M\"unchen, Germany}
	E.~Garcia,$^6$ 
	A.~Gillitzer,$^{1}$
	%%\footnote{Institut f\"ur Kernphysik, 
	%%Forschungzentrum J\"ulich, D-52425 J\"ulich, Germany}
	G.~Heintzelman,$^{7}$
	%%\footnote{Brookhaven National Laboratory, Upton,
	%%NY 11973, USA}
	W.F.~Henning,$^{1}$
	%%\footnote{Gesellschaft f\"ur Schwerionenforschung,
        %%D-64291 Darmstadt, Germany}
	D.J.~Hofman,$^{1,5}$
	%%\footnote{University of Illinois at Chicago, 
	%%Chicago, IL 60607, USA}
	B.~Holzman,$^5$ 
	J.H.~Kang,$^9$ 
	E.J.~Kim,$^9$
	S.Y.~Kim,$^9$ 
	Y.~Kwon,$^9$ 
	D.~McLeod,$^5$
	A.C.~Mignerey,$^6$ 
	M.~Moulson,$^{4}$
	%%\footnote{Laboratori Nazionali di Frascati, INFN,
	%%00044 Frascati, Italy}
	V.~Nanal,$^{1,5}$
	%%\footnote{Tata Institute of Fundamental Research,
	%%Colaba, Mumbai 400005, India}
	C.A.~Ogilvie,$^7$
	R.~Pak,$^8$
	A.~Ruangma,$^6$
	D.E.~Russ,$^6$ 
	R.K.~Seto,$^3$ 
	P.J.~Stanskas,$^6$
	G.S.F.~Stephans,$^7$ 
	H.Q.~Wang,$^3$
	F.L.H.~Wolfs,$^8$ 
	A.H.~Wuosmaa,$^1$ 
	H.~Xiang,$^3$
	G.H.~Xu,$^{3}$
	%%\footnote{University of Houston, Houston, TX 77004, USA}
	H.B.~Yao,$^7$ 
	and C.M.~Zou$^3$\\
	(E917 Collaboration)\\*[0.3cm]}
\address{
	$^1$Argonne National Laboratory, Argonne, IL 60439, USA\\
	$^2$Brookhaven National Laboratory, Upton, NY 11973, USA\\
	$^3$University of California, Riverside, CA 92521, USA\\
	$^4$Columbia University, Nevis Laboratories, Irvington, NY 10533, USA\\
	$^5$University of Illinois at Chicago, Chicago, IL 60607, USA\\
	$^6$University of Maryland, College Park, MD 20742, USA\\ 
	$^7$Massachusetts Institute of Technology, Cambridge, MA 02139, USA\\	
	$^8$University of Rochester, Rochester, NY 14627, USA\\
	$^9$Yonsei University, Seoul 120-749, South Korea}

\maketitle

% ----	ABSTRACT

\begin{abstract}

An excitation function of proton rapidity distributions 
for different centralities is reported from AGS Experiment E917 
for Au+Au collisions at 6, 8, and 10.8 GeV/nucleon.
The rapidity distributions from peripheral collisions have a valley 
at mid-rapidity which smoothly change to distributions that display
a broad peak at 
mid-rapidity for central collisions. 
The mean rapidity loss increases with increasing beam
energy, whereas the fraction of protons consistent with isotropic
emission from a stationary source at midrapidity decreases with
increasing beam energy. The data suggest that the stopping is substantially
less than complete at these energies.

\end{abstract}

\vspace{0.5cm}

Nuclear matter is believed to be compressed to high baryon density
($\rho_{B}$)
during central collisions of heavy nuclei at relativistic
energies\cite{Vide,Akib,PP}. 
In the interaction region of the colliding nuclei, nucleons 
undergo collisions which reduce their original longitudinal momentum. 
This loss of rapidity is an important characteristic of the reaction mechanism, 
and is often referred to as stopping\cite{Busza}.
The rapidity loss for beam nucleons has been extensively 
studied in p+A reactions\cite{Ledoux}
and more recently in heavy-ion reactions\cite{VandH,Harr,Hong,Appel,Barrette}.
Relativistic heavy-ion collisions are unique in the sense that
secondary collisions of excited baryons are expected to
contribute to the rapidity loss leading to simultaneous stopping 
of many nucleons within the interaction volume. 
The $\rho_{B}$ thus reached may be large enough to induce
phase transitions, such as quark deconfinement 
and/or chiral symmetry restoration \cite{Shuryak}.
The observation of large numbers of baryons at mid-rapidity is
indicative of compression to high $\rho_{B}$, but the
quantitative connection to $\rho_{B}$ is only possible via model
calculations. An important key
to our understanding of this phenomenon is systematic 
data for the rapidity
distributions of baryons over a broad range of conditions, 
such as beam energy and centrality.

In this Letter, we present an excitation function of the centrality
dependence of proton rapidity distributions from Au+Au
collisions at 6, 8, and 10.8 GeV/nucleon.
At these energies, the measured ratio of antiprotons to protons is
very small ($<0.03\%$)\cite{GH} and the production of protons from 
$\Lambda$ decay contributes less than 5\% to the total yield\cite{WE}.
The measured protons can therefore be considered to directly reflect
the distribution in rapidity of the initial  baryons (assuming that the
neutron rapidity distribution has a similar shape).
The measured rapidity distributions at all three beam energies show a
similar evolution with centrality. They are clearly bimodal in shape for
peripheral collisions and change to shapes which for central collisions may
still be bimodal in nature. This suggests that the degree of stopping
in central collisions is not complete, even in this heavy system. 
In addition, it is shown that the protons do not end up as an
isotropically emitting source at mid-rapidity, but retain a fair 
degree of their initial longitudinal motion.
This phenomenon is represented by a parameterized fit to the data with 
a superposition of longitudinally moving sources.

Experiment E917 measured Au+Au reactions at beam kinetic energies
of 6, 8, and 10.8 GeV/nucleon at the Brookhaven AGS, 
and was the final experiment in the 
series E802/E859/E866/E917\cite{Abbott,Kaon}.
The experimental apparatus in E917 consisted of a series of beam-line 
detector arrays which were used for global event characterization, and a
large rotatable magnetic spectrometer used to track and identify
particles.
The tracking system of the spectrometer consisted of a series
of drift and multi-wire ionization chambers, which bracketed either side
of a dipole magnet, followed by a segmented time-of-flight
wall of vertical scintillator slats.
The data presented here were taken with a trigger that required
at least one track in the spectrometer.
The centrality of an event can be selected through either of two 
approximately equivalent methods:
(1) from the multiplicity measured by a large acceptance device called
the New Multiplicity Array\cite{cuts}, or (2) from the energy 
deposited in the zero degree calorimeter\cite{cuts}. 
The data were sorted off-line into different centrality classes, where
normalization was provided by prescaled interaction triggers.
For each beam energy, the five event classes are reported as the percentage 
of the total interaction cross section ($\sigma_{\rm int}$~=~6.8~b),
corresponding to (0-5)\%, (5-12)\%, (12-23)\%,
(23-39)\%, and (39-81)\% ((39-77)\% for 10.8 GeV/nucleon) 
of $\sigma_{\rm int}$.

The systematic uncertainty on the normalization of the measured 
invariant spectra and rapidity distributions is dominated by
the uncertainty of the single-track efficiency and the loss 
of tracks due to hit-blocking.
The tracking uncertainty increases from 5\% to 10\% towards
mid-rapidity.
For peripheral collisions, there is a 10\% uncertainty in the 
cross section, which decreases to 5\% for central collisions.
These uncertainties lead to a total systematic uncertainty 
of 15\% independent of centrality, with a 5\% relative uncertainty
across beam energies. 

\begin{figure}[hbt]
\centerline{
\epsfig{file=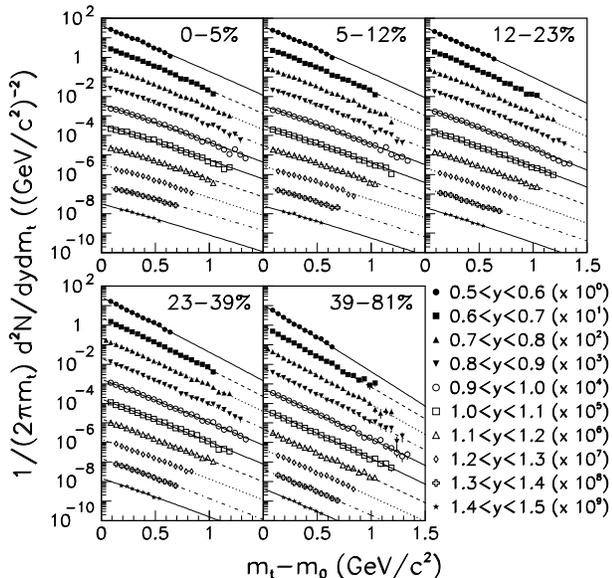,angle=-90,width=8cm}}
\caption[]{Invariant yields of protons as a function of transverse mass for
ten rapidity intervals for each centrality class of Au+Au collisions at 8
GeV/nucleon. The most backward rapidity in each panel is plotted on the
correct scale, while successive spectra have been divided by ten for
clarity. The errors are statistical only. The curves are Boltzmann fits
described in the text.}
\label{stopping_fig1}
\end{figure}

The measured invariant yields of protons from Au+Au collisions at 
8 GeV/nucleon are shown in ten rapidity intervals in Fig. \ref{stopping_fig1}
as a function of the transverse mass, $m_t=\sqrt{p_t^2+m_0^2}$. The spectra at 8 GeV/nucleon are representative of the
quality of the data at other beam energies. The errors are statistical only. 
The data were fit with a Boltzmann-form function in $m_{t}$:

\begin{eqnarray}
\frac{1}{2 \pi m_{t}} \frac{d^{2}N}{dm_{t}dy} =
\frac{dN/dy}{2 \pi (m_{0}^{2}T+2m_{0}T^{2}+2T^{3})}
m_{t}e^{-(m_{t}-m_{0})/T} 
\end{eqnarray}
where $T$ and $dN/dy$ are free parameters.
The data points were weighted according to their statistical errors.
As has been previously reported, for Au+Au collisions at
10.8 GeV/nucleon\cite{PP}, the proton spectra cannot be satisfactorily
described by a single exponential function.
In contrast, the above form reproduces the spectra well with
$\chi^2$ per d.o.f in the range 0.5 - 2.0, and provides 
the inverse slope parameter ($T$) and the rapidity density ($dN/dy$) 
for each rapidity interval.
  
\begin{figure}[hbt]
\centerline{
\epsfig{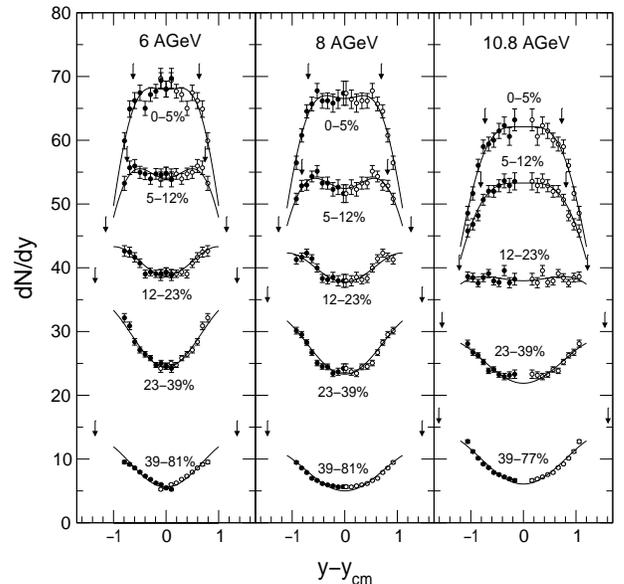}}
\caption[]{Proton rapidity distributions
for all centrality classes at all three
beam energies from Boltzmann fits to the invariant cross sections.
The open symbols are the data reflected about mid-rapidity.
The errors are statistical only. The curves represent double Gaussian fits
to the data (see text), the centroids of which are indicated by the arrows.}
\label{stopping_fig2}
\end{figure}

The centrality dependence of the rapidity density at each beam energy
is shown in Fig.~\ref{stopping_fig2}.
The data are shown relative to the rapidity of the center-of-mass
of the system ($y_{\rm cm}$) which equals 1.35, 1.47, and 1.61 for 
beam kinetic energies of 6, 8, and 10.8 GeV/nucleon, respectively.
The data points reflected about mid-rapidity are shown as open symbols.
The errors were calculated from the fitting procedure.
The present values of $dN/dy$ for the most central event 
class at 10.8~GeV/nucleon are in good
agreement with the previously reported values
from the E866 Collaboration\cite{PP}. 

\begin{figure}[hbt]
\center{
\epsfig{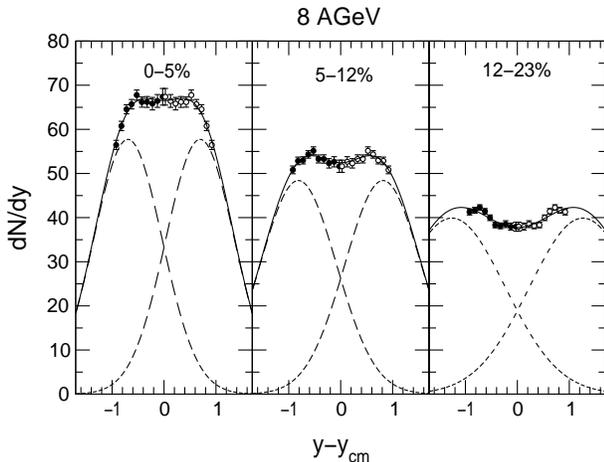}}
\caption{Fits to the proton $dN/dy$ distributions for the 
three most central event classes at 8 GeV/nucleon illustrating the evolution
of shape with centrality. The solid curves are the sums of two individual
Gaussians (dashed curves).}
\label{stopping_fig3}
\end{figure}

For each beam energy, a common trend is observed in the evolution 
of the shape of the rapidity distributions as a function of centrality.
For the most peripheral event class the distribution 
has a minimum value of $dN/dy~\approx$~6 at mid-rapidity.
This concave shape persists to the next most central event class,
corresponding to a centrality cut of (23-39)\%,
consistent with the expectation that most of the participant protons 
reside at beam 
rapidities following these relatively peripheral collisions.  
For more central collisions, the rapidity distributions become progressively 
flatter, and begin to develop a broad maximum at mid-rapidity 
for all beam energies. The trend of the data suggests, however, that for the
most central collisions, the distribution does not evolve to a
single peak centered at mid-rapidity, but rather is consistent with two 
components each displaced from mid-rapidity, or with a set of sources spread 
throughout the rapidity range. To qualitatively emphasize this point, we
have fit each of the measured distributions with two Gaussian peaks centered
symmetrically about mid-rapidity. These are shown superimposed on the  
data for the three most central event classes at 8 GeV/nucleon
in Fig.~\ref{stopping_fig3} and clearly display the trend. This result, 
somewhat surprisingly, implies that complete stopping is not
achieved at AGS energies and that the longitudinal rapidity distribution is
a result of transparency.

To quantify these results, we calculate the mean rapidity loss from the fits
for the most central collisions $\langle\delta y\rangle = y_{beam}-\langle y
\rangle$ where $\langle y \rangle$ refers to the Gaussians centered at
positive rapidity. These values are listed in Table I. To compare with
previous results for the most central data at 10.8 GeV/nucleon, we have also
calculated the mean rapidity loss for protons in the restricted rapidity
range $-1.11<y-y_{\rm cm}<0$ and find a value of 1.07$\pm$0.05, which is in good
agreement with the value of 1.02$\pm$0.01 obtained for the (0-4)\% most
central Au+Au collisions at the same beam energy reported by Videb{\ae}k and
Hansen \cite{VandH}. 

\begin{figure}[hbt]
\center{
\epsfig{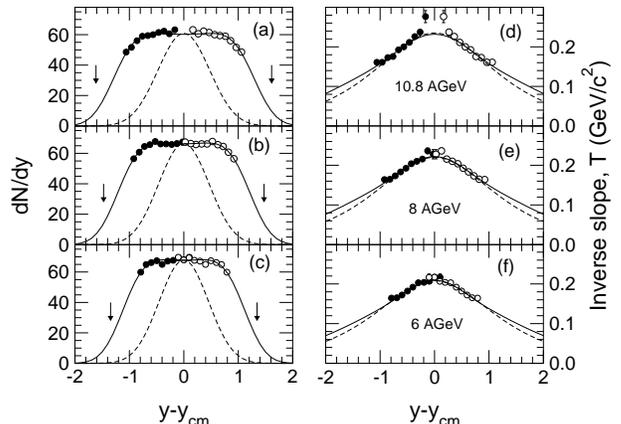}}
\caption{
Measured (solid points) and reflected (open circles)
proton rapidity distributions and inverse slopes are shown
as a function of 
rapidity $y-y_{\rm cm}$
for beam energies of 6, 8, and 10.8 GeV/nucleon. 
The arrows indicate the target and beam rapidities. 
The dashed curves represent 
the expected distribution for
isotropic emission from a thermal source at rest in the 
center-of-mass system ($y = y_{\rm cm}$), whereas the solid curves correspond 
to an optimum fit to the data for a uniform distribution of sources within 
a range of rapidities 
($y_{\rm cm}-y_b) < y < (y_{\rm cm}+y_b$)
with a Gaussian profile of
$T_{\rm eff}(y)$ centered at $y_{\rm cm}$ (see text). 
The parameters obtained from a least squares fit to the data are $y_b$~=~0.990, 
1.086, 1.166, $T_{\rm eff}^0$~=~0.253, 0.267, 0.279 GeV/$c^2$, and 
$\sigma_T$= 0.697, 0.762, 0.809 for 
$E_{\rm beam}$ = 6, 8, 10.8 GeV/nucleon, respectively.}
\label{stopping_fig4}
\end{figure}

The mean rapidity losses determined from our data show a systematic
increase with incident energy. The mean rapidity loss does not, however,
fully characterize all the features of the final rapidity distribution.
Despite the fact that a considerable amount of the original longitudinal
momentum of the incident particles has been lost, the situation is clearly
far from one where they are completely stopped. Fig.~\ref{stopping_fig4} 
shows
the $dN/dy$ distributions and inverse slopes for the (0-5)\% most central
event classes at all three energies plotted as solid points. The dashed
curves in Fig.~\ref{stopping_fig4} show the expected distribution for complete stopping - 
{\it i.e.} isotropic emission from a source at rest in the center-of-mass 
system emitting protons with a Boltzmann energy distribution, the effective
temperature of which is adjusted to reproduce the inverse slope of the 
transverse
mass spectrum at mid-rapidity. The effective temperature thus accounts for the
combined effects of radial expansion and temperature of this source.
Whereas this gives a good description of the rapidity dependence
of the inverse slope parameters, it completely fails for the $dN/dy$
distributions. In fact, only a fraction of the observed yields can be
accounted for in such a scenario. The yields can, however, be adequately
accounted for by emission from a continuum of isotropic sources,
uniformly distributed over a rapidity range $y_{cm}\pm y_b$. In order to
account for the rapidity dependence of the inverse slope parameter it is,
however, necessary to introduce a Gaussian rapidity dependency of the
effective temperature 
$T_{\rm eff}(y) = T_{\rm eff}^0\exp(-(y-y_{\rm cm})^2/2\sigma_{T}^2)$. 
The solid curves shown in Fig.~\ref{stopping_fig4} were
obtained by a four parameter simultaneous fit of
$y_b; T_{\rm eff}^0; \sigma_T$; and $N_0$
($dN/dy$ normalization) to the experimental $dN/dy$ and
inverse slope $T(y)$ values.
Based on the fit $dN/dy$-distributions, the total number of protons
within the range $-2<y-y_{cm}<2$ ($\int_{-2}^{2} \frac{dN}{dy} dy$) 
was found to be 155, 164, and 159 for $E_{\rm beam}$ = 6, 8, and 10.8 
GeV/nucleon, respectively. Since the total initial number of protons (158) is
approximately accounted for, this suggests that the extrapolation into the
unmeasured region is quite reasonable.

The fraction of the observed protons corresponding to complete stopping
($f_{iso}$) was calculated as the ratio of the areas under the dashed and
solid curves, respectively. The values of $f_{iso}$ at each energy are
listed in Table I.
The values of the absolute rapidity loss from the distributed source fits
are also listed in Table I. They are in close agreement with the values from
the Gaussian fits and show the same systematic increase with increasing
incident energy. This result is different from observations in p+A
collisions in which the rapidity loss was found to be independent of beam
energy \cite{Ledoux}. This would suggest an increased role of secondary
interactions which increase stopping in heavy-ion collisions. In contrast,
the fractional rapidity loss, $\langle \delta y \rangle/\delta y_{max}$,
where $\delta y_{max}=y_{beam}-y_{cm}$, is found to be essentially 
constant with beam
energy from 6 to 158 GeV/nucleon, similar to the trend noted in Ref.
\cite{Harr}.

In summary and conclusion, we have measured proton rapidity distributions
for Au+Au collisions at three energies as a function of collision
centrality. These distributions show a consistent evolution with increasing
energy and centrality leading to maximal rapidity loss for the most central
collisions and the highest energy. This result suggests the importance of
secondary reactions in the heavy-ion case, contrary to the situation in p+A
collisions. Nevertheless, the degree of stopping of the incident baryons is
far from complete and only a fraction of the observed protons can be
accounted for by the emission from a stopped isotropically emitting source.
The remainder still possess considerable longitudinal momentum.
It is not possible to say unequivocally from the present data alone whether 
or not this corresponds to a situation where the colliding nuclei are, to
some extent, ``transparent'', or a fully stopped and compressed system has
re-expanded. However, the systematic behavior of the centrality dependence
of the shapes of the rapidity distributions suggests, surprisingly, that the
former is the case.

This work is supported by the U.S. Department of Energy under
contracts with ANL (No.\ W-31-109-ENG-38), BNL (No.\ DE-AC02-98CH10886), MIT
(No.\ DE-AC02-76ER03069), UC Riverside (No.\ DE-FG03-86ER40271), UIC (No.\
DE-FG02-94ER40865), and the University of Maryland (No.\ DE-FG02-93ER40802),
the National Science Foundation under contract with the University of
Rochester (No.\ PHY-9722606), and the Ministry of Education and KOSEF (No.\
951-0202-032-2) in Korea. \\

% ----	Bibliography

% ----  Tables

\begin{table}[hbt]
\caption[] {Comparison between different measures of stopping in central
heavy-ion collisions. The absolute $\langle \delta y \rangle$ and relative
$\langle \delta y \rangle  / \delta y_{\rm max}$ rapidity losses
obtained from double Gaussian fits to the $dN/dy$-distributions
are given in column 2 and 3, respectively. The absolute and relative rapidity
losses obtained from the fitted solid curves in Fig.~\ref{stopping_fig4}(a-c) 
are given in column
4 and 5, respectively. Upper limits on the isotropic fraction $f_{\rm iso}$
of the total $dN/dy$ distribution (solid curves in 
Fig.~\ref{stopping_fig4}(a-c)) are given in
column 6. The last row lists parameters obtained from an analysis of 158
AGeV Pb+Pb collisions at 0-5\% centrality \cite{Appel}.}

\begin{tabular}{ccccccc}
$E_{\rm beam}$ &
\multicolumn{2}{c}{Double Gaussian} &
\multicolumn{2}{c}{Distributed sources} &
$f_{\rm iso}^{(a)}$\\
($A$GeV)  &
$\langle \delta y \rangle $ &
$\frac{\langle \delta y \rangle}{\delta y_{\rm max}}$    &
$ \langle \delta y \rangle^{(b)}$ &
$\frac{\langle \delta y \rangle}{\delta y_{\rm max}}$  & \\
\tableline
6.0   & 0.72$\pm$0.01 & 0.53 & 
        0.74$\pm$0.01 & 0.55 & 0.49$\pm$0.01\\
8.0   & 0.78$\pm$0.01 & 0.53 & 
        0.82$\pm$0.01 & 0.56 & 0.46$\pm$0.01\\
10.8  & 0.88$\pm$0.01 & 0.55 &
        0.93$\pm$0.01 & 0.57 & 0.45$\pm$0.01\\
158   & 1.71$\pm$0.02$^{(c)}$ & 0.59 & 
       -                      & -         & 0.23$\pm$0.02$^{(d)}$
\end{tabular}
\footnotesize{
$^a$Fraction of isotropically emitted proton (dashed curves in 
Fig.~\ref{stopping_fig4}(a-c)) of
total number of protons (solid curves of Fig.~\ref{stopping_fig4}(a-c)).\\
$^b$Statistical errors only.\\
$^c$Obtained from a double Gaussian fit, $dN/dy =
85[\exp(-(y-y_{\rm cm}+1.19)^2/1.6)+\exp(-(y-y_{\rm cm}-1.19)^2/1.6)]$
to the experimental data \cite{Appel}.\\
$^d$Obtained using $T_{\rm iso} = 0.26$ GeV/$c^2$, which reproduces
$\langle p_t \rangle $ at $y-y_{\rm cm}=0$ \cite{Appel}.}
\label{table:bb_tab}
\end{table}

% ----	Figures

\end{document}